# The FCC-ee design study: luminosity and beam polarization

**Michael Koratzinos**[1]

*University of Geneva, Switzerland*
*E-mail:* `m.koratzinos@cern.ch`

The FCC-ee accelerator is considered within the FCC design study as a possible first step towards the ultimate goal of a 100 TeV hadron collider. It is a high luminosity $e^+e^-$ storage ring collider, designed to cover energies of around 90, 160, 240 and 350GeV ECM (for the Z peak, the WW threshold, the ZH and ttbar cross-section maxima respectively) leading to different operating modes. We report on the current status of the design study, on the most promising concepts and relevant challenges. The expected luminosity performance at all energies, and first studies on transverse polarization for beam energy calibrations will be presented.

---

[1] Poster, on behalf of the FCC Design Study Group

**1. Introduction**

The discovery [1] [2] of the Higgs boson, the last missing link of the Standard Model (SM), and the realization that it is rather light (with a mass of around 125GeV) has renewed the interest of circular e+e- colliders that can serve as Higgs factories. Circular colliders enjoy several advantages compared to equivalent linear collider projects: Higher luminosity than a linear machine around the centre of mass energy ($E_{CM}$) needed for Higgs production via the Higgstrahlung process e+e-→ZH; multiple interaction points; mature technology that carries little risk; and finally the ability to run with very high luminosities at lower energies. They also suffer disadvantages the highest of which is their limited energy reach – a 80-100 km machine can reach 350GeV $E_{CM}$, where the couplings of the Higgs to the top quark could be directly measured, but cannot go much higher. Whether the superior energy reach of linear machines or the superior luminosity of circular machines is a decisive factor would critically depend on the physics landscape after the first results of the 13 TeV run of the LHC are digested.

The FCC design study, kicked off in 2014, addresses the possibility of a new circular tunnel in the Geneva region of 80-100 km circumference that will eventually house a 100 TeV proton collider. A possible intermediate step is a circular e+e- collider that can operate with excellent luminosities from $E_{CM}$ of about 90 GeV to 350 GeV, referred to as the FCC-ee. This paper will concentrate on the FCC-ee study. Already a very rich physics programme is under study which shows that such a machine can contribute decisively to the ultimate testing of the Standard Model and have the highest chance of giving us hints of what lies beyond [3].

**2. Maximum luminosity of a circular collider**

The luminosity of a circular electron collider with flat beams is given by the formula

$$\mathcal{L} = \frac{3}{8\pi} \frac{e^4}{r_e^4} P_{tot} \; \frac{\rho}{E_0^3} \xi_y \frac{R_{hg}}{\beta_y^*} \qquad (1)$$

where $r_e$ and $e$ are the classical radius of the electron and its charge, $P_{tot}$ the total SR power dissipated by one beam, $\rho$ the bending radius, $E_0$ the beam energy, $\xi_y$ vertical beam-beam parameter, $\beta_y^*$ the vertical beta function at the interaction point and $R_{hg}$ the geometric hourglass factor which is a function of $\beta_y^*$ and $\sigma_z$, the bunch length.

We can already note here that the luninosity depends linearly on the power dissipated and on the bending radius of the collider ( the dependence of luminosity to the bending radius gets modified slightly from a linear one due to the fact that the maximum achievable $\xi_y$ depends on damping and therefore on the machine radius for beam-beam dominated machines, see below). Luminosity also drops rapidly with increasing beam energy to the third power (again the relationship being modified slightly due to the dependence of $\xi_y$ to energy)

The vertical beam-beam parameter $\xi_y$ is defined as

$$\xi_y = \frac{N_b r_e \beta_y^*}{2\pi\gamma\sigma_x\sigma_y} \qquad (2)$$

where $\sigma_x, \sigma_y$ is the beam size in x and y and $N_b$ is the number of electrons (positrons) per bunch.

In case of non-head-on collisions, $\sigma_x$ and $\sigma_z$ are replaced by their effective lenghts. If the crossing angle of the beam is $\theta$ in the x-y plane then we define the Piwinski angle $\varphi$ as

$$\varphi = \frac{\sigma_z}{\sigma_x} \tan \frac{\theta}{2} \quad (3)$$

And $\sigma_x$ and $\sigma_z$ become

$$\sigma_{x,eff} = \sigma_x \sqrt{1 + \varphi^2} \quad (4)$$

and

$$\sigma_{z,eff} = \frac{\sigma_z}{\sqrt{1 + \varphi^2}} \quad (5)$$

The maximum achievable luminosity of a collider of a given power and radius in this formulation depends on the maximum value of the beam-beam parameter that can be achieved. This depends on if the machine is beam-beam dominated or beamstrahlung dominated [4].

For a reliable estimation of the maximum luminosity of a specific implementation, an approach that simulates dynamic effects is mandatory. If one ignores such effects, a luminosity estimate from the above formulas can be accurate to about 20%, provided that the beam interactions are not very disruptive (for example, in head-on collisions where $\sigma_z$ is smaller than $\beta_y^*$ and the beam-beam parameter values are not too close to the limit of eqn. (8) below).

## 2.1 The beam-beam limit

The beam-beam limit depends on the damping decrement, $\lambda_d$, the amount of energy loss when electrons move from one IP to the next:

$$\lambda_d = \left(\frac{U_0}{E_0}\right) \frac{1}{n_{IP}} \quad (6)$$

where $U_0$ is the energy loss per electron in one turn and $n_{IP}$ the number of interaction point regions. Using the following formulation from [5]

$$\xi_y^{max} \propto \lambda_d^{0.4} \quad (7)$$

we can fit the maximum beam-beam parameters achieved at LEP to derive the approximate formula for head-on collisions:

$$\xi_y^{max} \approx 0.86 \cdot \lambda_d^{0.4} \quad (8)$$

Please note that the above formulation is only based on a limited amount of LEP data and should be taken with a grain of salt. Beam-beam simulations and ultimately measurements on a real machine would provide a more accurate estimation.

## 2.2 The beamstrahlung limit

The beamstrahlung limit [6] is due to the fact that at high energies and luminosities, beamstrahlung, the synchrotron radiation emitted by an incoming electron in the collective electromagnetic field of the opposite bunch at an interaction point, reduces beam lifetimes to values where the top-up injector cannot cope. The effect of beamstrahlung is very implementation specific and can be mitigated by small vertical emittance and large momentum acceptance.

Two analytical calculations exist for computing beam lifetimes due to beamstrahlung [6] [7] offering fast estimates of the effect. Analytical simulations assume Gaussian distributions (i.e. without non-Gaussian tails) and have other approximations. Therefore it is important to be checked against a complete simulation such as the one in [8]. The comparison between the two analytical calculations and the simulation show reasonable agreement for momentum acceptances of interest here (between 1.5% and 2%) at beam energies of both 120GeV and 175GeV [9]. This justifies the use of the analytic formulas instead of the much more accurate but time-consuming simulation for the purposes of this work.

The two regimes (the beam-beam dominated and the beamstrahlung dominated one), for the specific implementation of FCC in [10] and for head-on collisions, can be seen in Figure 1. Such a machine would be beamstrahlung dominated at 175GeV and beam-beam dominated at 120 GeV.

Moving to a crab waist scheme allows for the beam-beam limit to be increased, therefore such a machine can run at the beamstrahlung limit even at low energies, largely increasing luminosity.

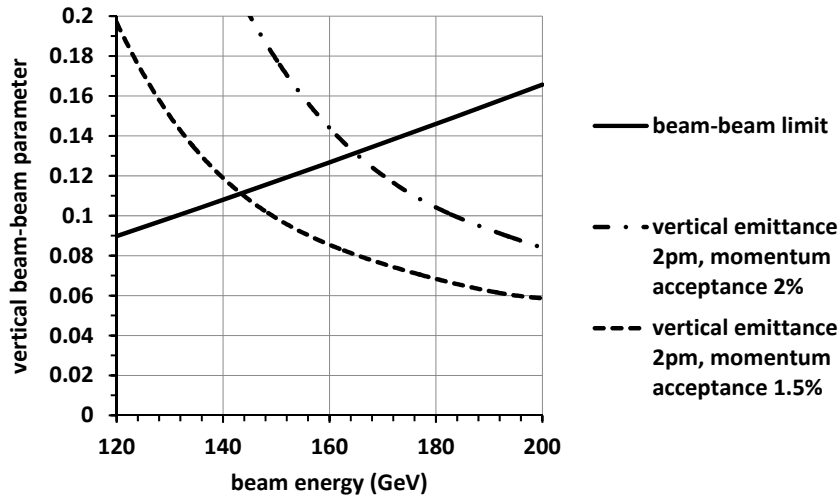

Figure 1: The beam-beam and beamstrahlung limits for a collider with parameters as in [10] (head-on collisions). The beamstrahlung curves have been obtained assuming a beam lifetime of 300 seconds and two different momentum acceptance values of 1.5% and 2%. The allowed region is the area below the beam-beam and beamstrahlung lines.

**2.3 Head-on versus colliding at an angle**

Colliding at a (small) crossing angle compared to colliding head-on offers a series of advantages and disadvantages. Some of the advantages are: fewer optics elements close to the IP, since the accelerator has two beam pipes in the arcs; no parasitic collisions; possibility to use the Crab Waist scheme and increase performance. And the disadvantages: crossing the solenoidal magnetic field of the experiments at an angle blows up the vertical emittance and needs to be compensated; and the design of the final focus quadrupoles is more demanding.

It should be noted that if no additional modifications are made, the luminosity performance of the two schemes is quite similar because the luminosity lost due to geometric effects is gained back due to the reduction of the beam-beam parameters. However, substantial gains can be achieved if the 'Crab Waist' scheme is implemented [11] [12]. This scheme introduces additional sextuples close to the interaction point to suppress betatron resonances and allow for much higher beam-beam parameters than otherwise possible. If the machine at a specific energy is bound by the beam-beam limit, then luminosity can be increased. If, however, the machine is already at the beamstrahlung limit, to first order no luminosity increase can be achieved. A machine like the FCC-ee with head-on collisions, as discussed, becomes beamstrahlung dominated at high energies, and remains beam-beam dominated at lower energies. Therefore a crab waist scheme with a crossing angle of 30mrad and crab sextuples gives a healthy luminosity increase at low energies (90 and 160 GeV $E_{CM}$), a small gain at 240GeV and no gain at 350GeV $E_{CM}$ [13]. The crab waist scheme is at the last stages of being tested with dynamic simulations and might well become the baseline scheme for FCC-ee. The preliminary luminosity performance of this scheme can be seen in Table 1. We should mention here that the luminosity figures shown are for colliding bunches of equal size. In practice there is always an asymmetry in the size of the electron and positron colliding bunches and this might lead to the so-called flip-flop instability. The amount of asymmetry above which the instability becomes important is under investigation and might lead to a reduction of achievable luminosity.

**Table 1:** Luminosity performance of the crab waist scheme per IP (preliminary, equal size colliding beams)

| Running mode ($E_{CM}$) | Z(90 GeV) | W(160 GeV) | H(240 GeV) | t(350 GeV) |
|---|---|---|---|---|
| Luminosity ($10^{34}$ cm$^{-2}$s$^{-1}$) | 210 | 38 | 8.7 | 2.1 |

## 3. Beam polarization

Accurate energy determination is a fundamental ingredient of precise electroweak measurements. In the case of LEP1 the centre of mass energy at and around the Z peak was known with an accuracy of around $2\times10^{-5}$. The exact contributions of the energy error to the mass and the width of the Z are presented in [14].

FCC-ee is capable of delivering statistics a factor $\sim10^5$ larger than LEP at the Z and WW energies, therefore there is a need not only to achieve similar performance as far as energy determination is concerned, but to do significantly better.

The beam energy of large storage rings continuously changes due to internal and extraneous causes. This evolution can be modelled, but energy changes are many orders of magnitude larger than the instantaneous accuracy of a depolarization measurement, as discussed below. For example, small changes in the diameter of the ring due to elastic deformations of the earth's crust (due to, for instance, tidal forces) can have a big effect on the energy of the electrons and positrons. This is due to the small momentum compaction factor $\alpha_c$ which relates changes in energy to changes in the orbit length of a storage ring:

$$\frac{\Delta E}{E} = -\frac{1}{\alpha_c}\frac{\Delta L}{L} \qquad (1)$$

where $L$ is the orbit length. A change of $4 \cdot 10^{-8}$ in the circumference of FCC-ee (typical for tide-induced changes) would modify the beam energy of 45 GeV by a very sizable 360 MeV.

The many other effects that contribute to energy changes are discussed in [15]. None of them has a very fast changing component, so monitoring the energy every ~10 minutes would ensure a negligible extrapolation error.

The RF configuration can give rise to different energies at the IPs and for electrons and positrons, as can the slightly different orbit for the separated rings, therefore both species should be measured, something that was not done at LEP.

The only method that can provide the accuracy needed is the so-called resonant depolarization technique [15], each measurement of which has an instantaneous accuracy of O(10$^{-6}$). It is based on the fact that the spin of an electron in a storage ring (in a perfectly planar machine and in the absence of solenoids) will precess $a\gamma$ times for one revolution, where $a$ is the anomalous magnetic moment and $\gamma$ the Lorenz factor of the electron and therefore the spin tune $\nu$ is

$$\nu = a\gamma = \frac{aE}{mc^2} = \frac{E[MeV]}{440.6486(1)[MeV]} \qquad (1)$$

Deviations from the above formula are small and are discussed in [16] and [17] where they were found negligible for LEP, but should be revised in view of the much improved precision aimed at the FCC-ee.

The average of all spin vectors in a bunch is defined as the polarization vector $\vec{P}$. Therefore the average energy of a bunch can be computed by selectively depolarizing a bunch of electrons or positrons which have been polarized to an adequate level and measuring the frequency at which this depolarization occurs. Beam polarization is usually measured by laser polarimeters which exploit the spin dependence of the Compton scattering cross section. The accuracy with which the instantaneous average energy of the bunch is computed using this method is O(100KeV) – a value much smaller than the beam energy spread.

### 3.1 Transverse polarization

Electron and positron beams in a storage ring naturally polarize due to the Sokolov-Ternov effect [18]. For the purposes of energy calibration, important figures of merit are the asymptotic value of polarization that can be reached and the time constant of polarization build-up.

The maximum achievable polarization value is given by the theory as

$$P_{max} = \frac{8}{5\sqrt{3}} \cong 0.924 \qquad (2)$$

however, numerous depolarizing effects (due to for instance machine imperfections) limit this number to lower levels.

For an initially unpolarised beam the time dependence for build up to equilibrium is

$$P(t) = P_{max}[1 - \exp(-t/\tau_{pol})] \qquad (1)$$

and the built up rate is (in natural units)

$$\tau_{pol}^{-1}[s^{-1}] \approx \frac{2\pi}{99} \frac{E[GeV]^5}{C[m]\rho[m]^2} \qquad (2)$$

where $C$ is the circumference of the storage ring and $\rho$ its bending radius. Therefore polarization times increase with the machine circumference and decrease with energy (**Table 2**). The use of wigglers [19] can decrease this time as discussed further.

**Table 2:** Polarization times without the help of wigglers in the absence of imperfections

| Storage ring | Circumference (km) | Bending radius (km) | E (GeV) | $\tau_{pol}$ (hours) |
|---|---|---|---|---|
| LEP | 27 | 3.1 | 45 | 5.8 |
| FCC-ee | 100 | 10 | 45 | 252 |
| FCC-ee | 100 | 10 | 80 | 16 |

### 3.2 Polarization and energy spread

One important limitation on achievable polarization levels comes from the energy spread of the beam. Energy spread scales approximately like

$$\sigma_E \propto \frac{E^2}{\sqrt{\rho}} \quad (3)$$

If we extrapolate from the measurements done at LEP [20] where the maximum energy where polarization was observed was 60.6GeV (at a level of around 8%) we get the values of **Table 3**. Polarization at the W pair threshold (80GeV) at FCC-ee seems possible. This is in contrast of what was achieved at LEP and another input to the physics case of this unique machine. Measurements in [20] also indicated that energy spreads larger than about 52MeV lead to a significant drop of polarization levels. Detailed simulations should eventually replace the empirical approach based on the LEP experience.

**Table 3:** Extrapolation of LEP data to other machines regarding the maximum energy below which polarization levels will be adequate for depolarization measurements

| Storage ring | C(km) | Maximum energy with polarization (GeV) |
|---|---|---|
| **LEP** | 27 | 61 |
| **FCC-ee** | 80 | 80 |
| **FCC-ee** | 100 | 84 |

### 3.3 Resonant depolarization at the FCC-ee

The way that resonant depolarization measurements are performed is the following: Only one bunch is targeted at a time. Since the colliding rate is much larger than the polarization rate, for polarization to build up, this bunch needs to be a non-colliding bunch. It should be stated here that operation with colliding and non-colliding bunches might be a challenge due to the different tune shifts of the two species of bunches. The measurement proper consists of measuring the spin precession frequency by introducing a resonance in a 'trial and error' fashion. If no depolarization is observed, the frequency used is not the correct depolarizing frequency. The bunch remains polarized. If the bunch depolarises, the frequency corresponds to the exact mean energy of the bunch at that moment. To observe the polarization change, polarization levels of 5-10% are needed depending on the polarimeter.

### 3.4 Wigglers

The natural polarization time for large rings is very long as seen in **Table 2** (even though we only need polarization levels of 5-10%, so that we can divide the numbers in the table by a factor 10 to 20). The expected mean time between failures at FCC-ee cannot be assumed to be more than a few hours or a day, the same order of magnitude as polarization times to 10%. A way to reduce polarization times is the use of wigglers [19]. Wigglers are dipole magnets with two parts: a low field region and a high field region so that the integral field seen by the electrons is zero. However they help, as polarization time scales with the square of the field and polarization levels are not affected provided that the wiggler asymmetry (the ratio of lengths of the positive and negative field magnets) is larger than ~5.

Wigglers have, however, two undesired effects: They increase the energy spread and they contribute to the SR power budget of the machine. Therefore a possible strategy would be to use them is such a way that the energy spread is less than some pre-determined maximum and to switch them on only where necessary.

The maximum energy spread that can be tolerated can be determined by simulation or, more pessimistically, by using the LEP experience where, as discussed earlier, was determined to be around 52MeV. In the absence of a new design, we consider the wigglers suggested for LEP [19] that have an asymmetry of 6.15 and pole lengths of 0.65m and 4m for the strong and the weak field respectively.

The polarization time and wiggler SR power dissipated for various configurations can be seen in **Table 4**. These results have been obtained by simulation (SLIM) and are close to the analytical calculation. In each case we have pushed the wiggler field while keeping the energy spread below 52MeV. B+ is the field of the strong pole. As can be seen, polarization times are reduced by a large factor when using wigglers. Interestingly, polarization times depend only weakly on the number of wigglers installed (but a higher field per wiggler is needed).

Therefore useful polarization levels (5-10%) are reached after 70-140 minutes. The SR power dissipated by the wigglers is rather large, although it is reduced if one operates one wiggler at a high field rather than many at a reduced field. It should be noted here that wigglers introduce more damping and might help to achieve higher beam-beam parameters, partly compensating the luminosity loss due to wiggler SR power – this is a topic that needs to be investigated.

### 3.5 Wiggler operation at the FCC-ee

A possible strategy for energy calibration therefore emerges: Wigglers are mandatory. For the case of FCC-ee, 250 non-colliding bunches are sufficient. The wigglers can be switched on as soon as the machine starts filling up and can be switched off when 5-10% polarization is achieved. Machine fill-up times are expected to be around 30 minutes, therefore an extra ~50-100 minute dead time is introduced while polarization builds up and during which period no meaningful energy measurement can be performed. Also, due to the power taken up by the wigglers, the luminosity of the machine will be lower than during normal operation. Physics studies which do not need precise energy determination can take place, though.

When the required level of polarization for the non-colliding bunches has been achieved, the wigglers can be turned off and the depolarization measurements can start. Measuring and replacing 5 bunches for 5 depolarization measurements per hour, the FCC-ee will exhaust all non-colliding bunches in 50 hours, during which time the used non-colliding bunches will have been polarized again to more than 10%. We will investigate if wiggler operation at a reduced setting during physics could be beneficial to the energy determination or overall performance. Also, the study of collimating the large amount of radiation from the wigglers will be a priority.

We here assume that the number of electrons in a non-colliding bunch would be similar to the number of electrons of a normal (colliding) bunch. For the FCC-ee this number is $\sim 1.8 \cdot 10^{11}$ (similar to the LEP1 value). Having 250 out of 16700 bunches not colliding leads to an inefficiency of 1.5%.

**Table 4:** The effect of the use of wigglers on polarization times, energy spread and wiggler power dissipation according to the SLIM simulation and for the wiggler design described in [19]. B+ is the magnetic field of the short (strong) dipole of the wiggler.

| Machine | Energy (GeV) | No. of wigglers | B+ (T) | $\tau_{pol}$ (hours) | $P_\infty$(%) | $\tau_{10\%}$ (hours) | Energy spread (MeV) | Wiggler SR power/beam (MW) |
|---|---|---|---|---|---|---|---|---|
| TLEP | 45 | 0 | 0 | 252 | 92.4 | 27.3 | 17 | 0 |
| TLEP | 45 | 12 | .62 | 24.1 | 88.1 | 2.7 | 50 | 15 |
| TLEP | 45 | 1 | 1.35 | 27.6 | 88.1 | 3.1 | 50 | 7 |

### 3.6 Simulation

Polarization is a strong function of machine misalignment and non-linear calculations are mandatory for evaluating the effect of the energy spread in presence of machine imperfections. Initial simulations are discussed in [21] where after orbit correction polarization levels have been restored to acceptable values. This work will continue but it is clear that well-planned state-of-the-art correction schemes will be needed to restore polarization level losses due to misalignments.

## 4. FCC-ee challenges

The FCC-ee study is now in full swing and first solid parameters are falling into place. The optics study is well under way as are the beam-beam simulations. A very challenging environment for the design is the interaction region. Horizontal bends close to the IP are necessary to correct chromaticity and achieve high luminosity performance, but the synchrotron radiation created by those bends needs to avoid shining on the experiment. Therefore the critical energy of the synchrotron radiation of those magnets needs to be a lot lower than in the arcs (aiming for values around 100keV whereas the arcs have ~1MeV at a beam energy of 175GeV) and the total power dissipated small compared to the loss in the arcs also.

Another challenging area is the magnet design in the vicinity of the interaction point. Experiments have powerful and large solenoid magnets and the beam arrives at a small angle (15mrad half-angle). Moreover, the very small $\beta_y^*$ values needed for high luminosity necessitate the final focusing quadrupoles to be very close to the IP (2m in the baseline design). These quadrupoles are close together (beam distance of 6cm on the IP side and 16.5cm at the other end), probably necessitating some kind of combined design. The crosstalk of one quadrupole to the other should also be compensated for. These quadrupoles need to be shielded from the magnetic field of the experimental solenoid by means of a compensating solenoid having the same field magnitude but opposite sign. Finally, between these elements and the IP needs to be another anti-solenoid so that the integrated solenoidal field seen by the beam is zero. If the length of this anti-solenoid is 1m then the magnitude of its field needs to be double than the magnetic field of the experimental solenoid. All these magnetic elements need to be designed, a work which is now underway.

## 5. Conclusions

The FCC-ee is a proposed lepton circular collider able to deliver high luminosities for a precise study of the Z, W and Higgs bosons as well as the top quark at $E_{CM}$ energies ranging from 90 to 350GeV. A new colliding scheme, the Crab waist scheme looks promising in delivering higher luminosities than our original calculations, exploiting the potential of circular machines to the limit.

The resonant depolarization method seems accessible at the Z (45GeV per beam) and W (80GeV per beam) energies. Non-colliding bunches are mandatory for the measurement. Both lepton species should be measured. Long polarization times necessitate the use of wigglers, which however are needed only during a short period at the beginning of a fill. Measurements should be performed routinely at a rate of a few per hour.

Challenges remain but ideas of how to overcome them are in place and work is underway.